\RequirePackage{fix-cm}
\documentclass[final]{svjour3}       
\smartqed  
\usepackage{appendix}
\usepackage{amsmath}
\usepackage{graphicx}
\usepackage{array}
\usepackage{longtable}
\usepackage{natbib}
\setcitestyle{aysep={}}

\usepackage[hidelinks]{hyperref}
\usepackage{amssymb}
\usepackage{xcolor}

\begin{document}

\title{On the departure from Monin--Obukhov surface similarity and transition to the convective mixed layer}

\titlerunning{Surface similarity and transition to the convective mixed layer}

\author{Michael Heisel \and Marcelo Chamecki}

\institute{M. Heisel \at
              School of Civil Engineering, The University of Sydney, Sydney, NSW 2008, Australia
              \email{michael.heisel@sydney.edu.au}           
           \and
           M. Chamecki \at
           Department of Atmospheric \& Oceanic Sciences, University of California Los Angeles, Los Angeles, CA 90095, USA
}

\date{Submitted: DD Month YEAR}

\maketitle

\begin{abstract}

Large-eddy simulations are used to evaluate mean profile similarity in the convective boundary layer (CBL). Particular care is taken regarding the grid sensitivity of the profiles and the mitigation of inertial oscillations in the simulation spin-up. The nondimensional gradients $\phi$ for wind speed and air temperature generally align with Monin--Obukhov similarity across cases but have a steeper slope than predicted within each profile. The same trend has been noted in several other recent studies. The Businger-Dyer relations are modified here with an exponential cutoff term to account for the decay in $\phi$ to first-order approximation, yielding improved similarity from approximately 0.05$z_i$ to above 0.3$z_i$, where $z_i$ is the CBL depth. The necessity for the exponential correction is attributed to an extended transition from surface scaling to zero gradient in the mixed layer, where the departure from Monin--Obukhov similarity may be negligible at the surface but becomes substantial well below the conventional surface layer height of 0.1$z_i$. 
 
\keywords{Surface layer \and Convective boundary layer \and Monin-Obukhov similarity \and Large-eddy simulation}
\end{abstract}

\section{Introduction}
\label{sec:intro}

Within the atmospheric boundary layer (ABL), the surface layer is unsurprisingly the region directly above the Earth's surface. This layer is often described in terms of its properties -- approximately constant flux, negligible rotation effects, and adherence to surface scaling -- rather than formally defined \citep{Sutton1953,Kaimal1994}. One common convention is to assume the surface layer extends to the lowest 10\% or so of the ABL \citep{Stull1988,Garratt1994}, consistent with the depth of the logarithmic (log) region in more general wall-bounded flows \citep{Pope2000}.

With respect to the surface scaling property, the scaling of mean flow statistics within the surface layer is given by Monin--Obukhov similarity theory \citep[MOST,][]{Monin1954,Foken2006} that extends the log law of the wall \citep{Karman1930,Prandtl1932,Millikan1938} to thermally stratified conditions. MOST predicts universal similarity for the nondimensional mean gradients:

\begin{align}
\frac{\partial U}{ \partial z}\left( \frac{\kappa z}{ u_*} \right) &= \phi_m \nonumber \\
\frac{\partial \theta}{ \partial z}\left( \frac{\kappa z}{ \theta_*} \right) &= \phi_h, \label{eq1}
\end{align}

\noindent where the gradients are normalized using the log law definition and functions for $\phi$ must be determined empirically. Here, $z$ is height above the surface, $U(z)$ is the mean horizontal wind speed, $\theta(z)$ is the mean potential temperature, $u_*$ is the surface friction velocity scale, $\theta_*$ is the surface temperature scale, and $\kappa$ is the von K\'{a}rm\'{a}n constant. The velocity and temperature scales are related to the surface momentum flux as $\overline{u^\prime w^\prime}_s=-u_*^2$ and surface heat flux as $\overline{w^\prime \theta^\prime}_s = - u_* \theta_*$ such that Eq. \eqref{eq1} is often referred to as the flux-gradient relations. Owing to the assumed absence of other length, velocity, or temperature scales in the theory, $\phi_m$ and $\phi_h$ are considered functions only of $\zeta = z/L$ defined using the \citet{Obukhov1946} length

\begin{equation}
L = \frac{u_*^2 \theta_s}{\kappa g \theta_*},
\label{eq2}
\end{equation}

\noindent where $\theta_s$ is the mean surface temperature and $g$ is the gravitational constant.

Following the introduction of MOST, evaluations of field measurements from meteorological towers have largely corroborated the surface layer theory and universality of $\phi_{m}(\zeta)$ and $\phi_{h}(\zeta)$. For a convective ABL (CBL) with $L<0$ typical of daytime conditions, several experimental campaigns and reevaluations proposed power-law relations for $\phi_m$ and $\phi_h$ with some variability in the fitted parameters but a consistent general form for the functions \citep[see, e.g.,][]{Dyer1970,Businger1971,Carl1973,Yaglom1977,Hogstrom1988,Wilson2001,Katul2011}. The most common of these empirical relations are the Businger-Dyer profiles for convective conditions \citep{Businger1971,Dyer1974}

\begin{align}
\phi_m &= (1-16 \zeta)^{-0.25} \nonumber \\
\phi_h &= (1-16 \zeta)^{-0.5}, \label{eq3}
\end{align}

\noindent where the values of the constants depend modestly on the analysis. Agreement of field measurements with MOST and the empirical relations can be further improved by accounting for additional effects not considered in the idealized theory, e.g. time-dependent variability from large-scale turbulence \citep{Salesky2020} and anisotropy due to complex conditions \citep{Stiperski2023}.

More recently, direct numerical simulations and large-eddy simulations (LES) of the CBL have produced consistent trends that support MOST to first-order approximation but also reveal possible shortcomings: gradient statistics align with the Businger-Dyer profiles when comparing results across different simulated conditions \citep{Maronga2017}, but the decay in $\phi(\zeta)$ within each individual profile is steeper than predicted by Eq. \eqref{eq3}, demonstrating a lack of universality in $\phi(\zeta)$ \citep{Khanna1997,Pirozzoli2017,Li2018}. Accordingly, it has been proposed that surface layer gradients -- particularly for velocity -- may additionally depend on the boundary layer depth due to the influence of large-scale motions from the well-mixed layer that forms the bulk of the CBL \citep{Khanna1997,Johansson2001}. This idea is indirectly supported by observed trends in field measurements that suggest a parameter space beyond $\zeta$ is required to account for variability in gradient statistics \citep{Salesky2012}.

The connection between turbulent eddies in the mixed layer and surface layer statistics has been substantiated by several analyses of turbulent structure in the CBL. Conditional statistics show that the steep decay in $\phi(\zeta)$ and deviations from MOST noted above are predominately associated with large-scale turbulent events such as downdrafts \citep{Li2018,Fodor2019}. In more general terms, these deviations are related to the modulation of near-surface turbulence by buoyancy-driven eddies from aloft \citep{Lemone1976,Smedman2007,Gao2016,Salesky2018,Liu2019,Dupont2022}. A signature of the boundary layer depth also appears in the velocity and temperature spectra within and above the surface layer \citep{McNaughton2007,Chowdhuri2019}.

There have been relatively few attempts to model the modulation of surface layer gradients. \citet{Santoso1998} modeled mean wind profiles with a power law and exponential decay to achieve a smooth transition from the surface to the uniform mixed layer. \citet{Gryning2007} combined surface scaling with a constant mixed layer length scale, but the foremost goal was to extend similarity to higher positions above the surface layer. \citet{Salesky2020} corrected Eq. \eqref{eq3} for local-in-time deviations due to large-scale fluctuations. \citet{Li2021} quantified the deviation as a nonlocal transport through the framework of eddy diffusivity models. \citet{Cheng2021} and \citet{Liu2022} both introduced a correction to $\phi$ prescribed as a function of $z_i/L$, where $z_i$ is the base height of the stable capping inversion and is typically used to define the CBL depth.

One cautionary note regarding the previous findings and models is that several of the studies used simulations confined to relatively low Reynolds number compared to the ABL \citep{Pirozzoli2017,Li2018,Fodor2019,Cheng2021}. Considering the log law only emerges for high Reynolds numbers \citep{Marusic2013,Sillero2013,Lee2015}, the results may reflect a combination of buoyancy effects and finite Reynolds number corrections to the log law. Additionally, for wall-modeled LES studies the grid convergence of surface layer statistics is often not scrutinized. The presence of these limitations precludes a careful quantitative comparison of deviations from MOST observed across the literature.

In the present work, recurring trends in $\phi(\zeta)$ observed from simulations are further evaluated using new LES of the idealized dry CBL. In consideration of the effects noted above, the LES is for the inviscid limit and includes a detailed test of grid sensitivity. To account for the observed trends, the dependencies of $\phi_m$ and $\phi_h$ in Eq. \eqref{eq1} are expanded to include $z_i$ following the suggestion of \citet{Khanna1997}. The approach also considers recent evidence of outer layer stratification influencing surface similarity under stable conditions \citep{Heisel2023}, except in this case the similarity is influenced by free convection in the mixed layer. The goal is to empirically explain the behavior of the simulated mean profile statistics in the broader context of the transition from the surface layer to free convection in the mixed layer, while also reconciling the simulation results with the widespread support for MOST from field experiments discussed above. The proposed explanation -- an approximately exponential decay in $\phi$ as a function of $z/z_i$ -- has a limited effect on statistics very close to the surface where many field measurements are acquired, is qualitatively consistent with the profile shapes seen in recent simulation studies, and accounts for the profile transition between the surface and mixed layers with reasonable accuracy. The remainder of the article is organized as follows: the new LES cases are described in Sect. \ref{sec:les}; mean profile similarity is assessed in Sect. \ref{sec:results}; implications for resistance laws in the mixed layer and for the definition of the surface layer are discussed in Sect. \ref{sec:discussion}; finally, a summary is given in Sect. \ref{sec:conclusion}.

\section{Large-eddy simulations}
\label{sec:les}

The present simulations were conducted using standard practices for representing an idealized dry convective ABL \citep[see, e.g.,][]{Deardorff1972,Moeng1994,Sullivan1994,Noh2003,Salesky2017}: a range of unstable conditions was achieved by imposing different combinations of fixed surface heat flux $Q_*=\overline{u^\prime w^\prime}_s$ and geostrophic wind speed $U_g$, and the boundary layer was confined by a stable capping inversion. The inversion was introduced through the initial temperature profile following \citet{Sullivan2011}, which includes a uniform temperature in the boundary layer, a strong lapse rate $\Gamma = $ 0.08 K\,m$^{-1}$ in the range $z=$ 1000 to 1100 meters, and a weaker lapse rate $\Gamma = $ 0.003 K\,m$^{-1}$ to form the top-most capping inversion.

Additional imposed parameters include the aerodynamic roughness length $z_o=$ 0.1 m and Coriolis frequency $f=1\times 10^{-4}$ s$^{-1}$. Six primary cases with varying $Q_*$ and $U_g$ and the resulting scaling parameters are summarized in Table \ref{tab1}. A seventh case (E1) is the same as case E, but with the initial capping inversion positioned 200 m lower as seen in the $z_i$ values that were determined from the height of the minimum heat flux. The range of simulated conditions span from relatively weak ($-z_i/L=$ 2.5) to moderately strong ($-z_i/L=$ 39) convection. All cases employed a numerical grid with 1024$\times$1024$\times$512 points and corresponding domain dimensions of 12$\times$12$\times$2 km.

\begin{table}
\caption{Key scaling parameters for large-eddy simulations (LES) of the convective atmospheric boundary layer (CBL) on a 1024$\times$1024$\times$512 numerical grid: imposed geostrophic wind speed $U_g$, imposed surface heat flux $Q_*$, friction velocity $u_*$, surface temperature scaling $\theta_*$, Obukhov length $L$, boundary layer depth based on the capping inversion height $z_i$, and bulk instability parameter $z_i/L$.}
\label{tab1}       
\begin{tabular}{ccccccccc}
\hline\noalign{\smallskip}
Case		& $U_g$	& $Q_*$	& $u_*$	& $-\theta_*$	& $-L$	& $z_i$	& $-z_i/L$	\\
		& (m\,s$^{-1}$)	&(K\,m\,s$^{-1}$)	& (m\,s$^{-1}$)	& (K)	& (m)	& (m)	& (--)	\\
\noalign{\smallskip}\hline\noalign{\smallskip}
A		& 15		& 0.1	& 0.81	& 0.12			& 415	& 1040	& 2.5		\\
B		& 15		& 0.17	& 0.82	& 0.21			& 256	& 1080	& 4.2		\\
C		& 15		& 0.24	& 0.83	& 0.29			& 188	& 1110	& 5.9		\\
D		& 12		& 0.24	& 0.71	& 0.34			& 119	& 1110	& 9.3		\\
E1		& 9		& 0.24	& 0.56	& 0.43			& 60		& 926	& 15.5		\\
E		& 9		& 0.24	& 0.58	& 0.41			& 65		& 1110	& 17.0		\\
F		& 6		& 0.24	& 0.44	& 0.54			& 29		& 1110	& 38.8		\\		
\noalign{\smallskip}\hline
\end{tabular}
\end{table}

\begin{figure}
\centering
  \includegraphics[width=\linewidth]{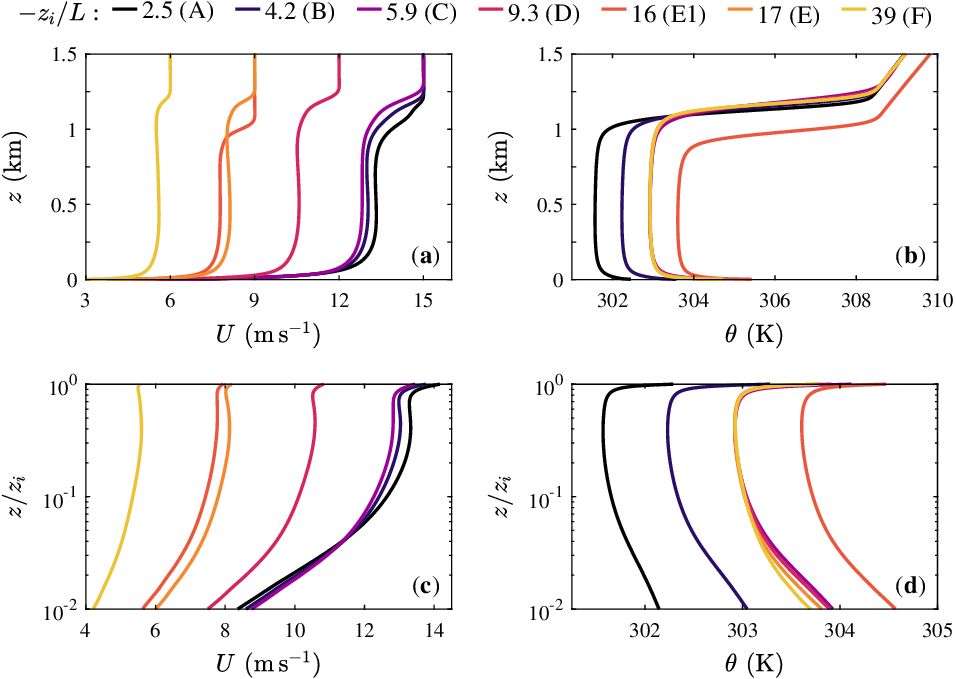}
\caption{Mean profiles for each LES case in Table \ref{tab1}. Columns correspond to the horizontal wind speed $U=(U_x^2 + U_y^2)^{1/2}$ (a,c) and potential temperature $\theta$ (b,d). Rows correspond to dimensional height (a,b) and relative height $z/z_i$ with logarithmic spacing (c,d) to show the near-surface trends. In this and later figures, the legend corresponds to the $z_i/L$ stability parameter and alphabetical label for the cases in Table \ref{tab1}. The temperature profiles in cases C, D, E, and F overlap in the outer layer due to the matching $Q_*$ and initial $z_i$.}
\label{fig1}     
\end{figure}

Dimensional profiles of the mean horizontal wind speed and potential temperature are respectively shown in Figs. \ref{fig1}a and \ref{fig1}b for the seven simulated cases. Every case exhibits an extensive mixed layer with approximately uniform wind speed and air temperature, as well as an entrainment layer centered around $z_i$ and an overlying temperature inversion. The temperature in the mixed layer increases with $Q_*$ and is highest for case E1 with a shallower CBL that can be heated more quickly. Hereafter, each case is referred to using its alphabetical label A-F indicated in the figure legends alongside the $z_i/L$ values. Unless otherwise noted, ``velocity'' refers to the magnitude of the horizontal components $U_x$ and $U_y$, and ``temperature'' refers to the potential temperature. Further details on the numerical code and procedure used to generate the Fig. \ref{fig1} profiles are given below.

The LES flow solver and numerics are a modified version of the code by Albertson and co-authors \citep{Albertson1996,Albertson1999,Porte2000}. The solver utilizes a staggered vertical grid for the vertical velocity component to optimize the vertical differentiation with second-order-accurate finite differencing. Horizontal derivatives are computed spectrally with full de-aliasing using zero padding. The flow is evolved through time integration using the second-order-accurate Adams-Bashforth method. The boundary conditions are periodic in the horizontal directions and zero penetration along the bottom and top of the domain, with the top additionally employing stress-free conditions. Fluctuations in the top 25\% of the domain are damped to achieve the upper boundary conditions and mitigate gravity waves \citep{Nieuwstadt1993}.  Subgrid-scale (SGS) stresses are estimated using a Lagrangian averaged scale-dependent model to represent the SGS eddy viscosity \citep{BouZeid2005}. The SGS heat flux is estimated using the same value as the local SGS eddy viscosity along with a constant Prandtl number $\mathrm{Pr}_{sgs}=$ 0.4. A more detailed description is given elsewhere \citep{Kumar2006,Salesky2017}.

The surface stress in the wall-modeled LES is estimated locally in space and time using MOST and Eq. \eqref{eq3} for momentum \citep{Moeng1984}. The wall model for $u_*$ is subject to so-called overshoot and log-law mismatch \citep{Mason1992,Brasseur2010,Larsson2015}. To mitigate this effect, the wall model was evaluated using velocity values at 0.05$z_i$ rather than the first grid point \citep{Kawai2012}, where $z_i$ in this case is based on the initial temperature profile and for simplicity does not change with time. Based on a comparative test, it was found that using values farther from the surface reduces the wall model overshoot but does not noticeably improve the convergence discussed below.

The final methodological considerations are the procedure to spin up the simulations from initial conditions and the sensitivity of the results to grid resolution. Both of these aspects are crucial to the repeatability and validity of the study and an extended discussion is given for each in the two subsections below. Statistical uncertainty is discussed later in Sect. \ref{sec:results} in the context of the results.

\subsection{Simulation spin-up and inertial oscillations}
\label{subsec:spinup}

A common approach for simulating the CBL is to impose an initially geostrophic velocity field with no boundary layer, i.e. with $U_x(z) = U_g$ and $U_y(z)=0$. The simulations are then spun up for a duration in the range of $t w_* / z_i \approx$ 5 to 30 turnover times until statistics appear stationary \citep[see, e.g.,][among others]{Moeng1994,Noh2003,Salesky2017,Maronga2017,Li2018,Liu2023}, where $z_i$ and the convective velocity $w_*$ \citep{Deardorff1970} represent properties of the large-scale eddies that govern dynamics in the bulk of the CBL. While the reasoning in this approach is sound, it neglects the presence of inertial oscillations resulting from the initial velocity field \citep[e.g.,][]{Shibuya2014,Momen2017}, noting the oscillations are not relevant to free convection cases with no geostrophic wind.

\begin{figure}
\centering
  \includegraphics[width=\linewidth]{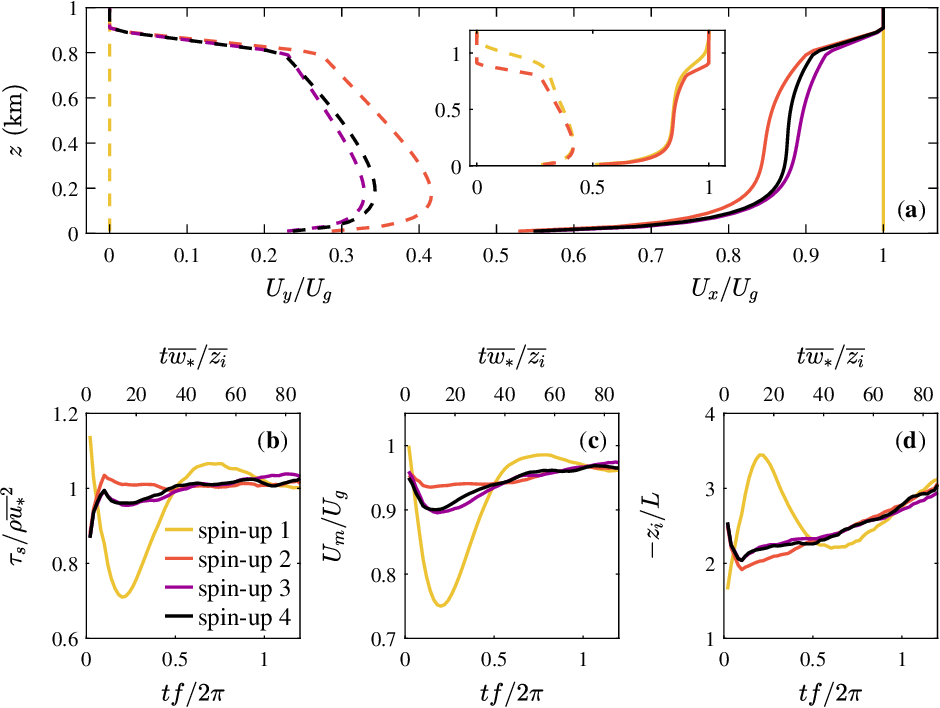}
\caption{Demonstration of the employed spin-up procedure and the impact of initial conditions on inertial oscillations. For the initial velocity profiles in (a), the resulting average surface shear stress $\tau_s$ (b), mixed layer wind speed $U_m$ (c), and stability $z_i/L$ (d) are plotted over time for one inertial period. During a given spin-up, the velocity profiles at $tf=\pi$ are rescaled to the original $z_i$ and used as the initial profiles for the subsequent spin-up, as shown in the inset of (a). Parameters with overbars ($\overline{\,\cdot\,}$) indicate the long-term average value over the full period.}
\label{fig2}     
\end{figure}

When the $U_x$ and $U_y$ fields differ significantly from the quasi-equilibrium condition, an oscillatory response arises from the Coriolis force and the oscillations dampen in time due to the surface stress \citep{Schroter2013}. An example is shown in Fig. \ref{fig2}, where the trivial initial profile in \ref{fig2}a (spin-up 1, yellow) results in an oscillating time evolution of average surface shear stress $\tau_s$ (\ref{fig2}b), mixed layer wind speed $U_m$ (\ref{fig2}c), and stability $z_i/L$ (\ref{fig2}d). The same time evolution occurs during spin-up of the conventionally neutral ABL \citep{Pedersen2014,Liu2021}.

For the present example, case A in Table \ref{tab1} is employed on a coarser 200$\times$200$
\times$100 numerical grid with $f=1.4\times 10^{-4}$ s$^{-1}$ increased to a typical polar value and the initial inversion height reduced by 200 m in the same manner as case E1. The latter two changes were necessary to prevent the CBL from reaching the damping layer before completing an inertial period 2$\pi/f$. Parameters with overbars ($\overline{\,\cdot\,}$) indicate the long-term average value over the full period.

Ending the simulations near $t w_* / z_i\approx$ 20 when the statistics are momentarily stationary would lead to shear stress and wind speed values that are significantly out of equilibrium with the conditions given by $Q_*$, $z_o$, $U_g$, $f$, and the initial $z_i$. Of particular importance to the present analysis is the Obukhov length $L$ which varies by approximately 50\% within the first half oscillation for the example in Fig. \ref{fig2}d.

The amplitude of inertial oscillations can be reduced by using initial $U_x$ and $U_y$ velocity fields that are closer to the quasi-equilibrium. To this end, an ad hoc procedure using multiple spin-up trials was developed to determine appropriate initial conditions for the final simulations. For the initial velocity profiles of the second spin-up, the mean conditions at $tf=\pi$ in the first spin-up were rescaled along $z$ to match the original $z_i$ as shown in the inset of Fig. \ref{fig2}a. Additionally, the original initial temperature profile \citep{Sullivan2011} was used rather than a rescaled temperature profile in order to restore the two initial inversion layers. The profile re-scaling and new spin-ups can be repeated as necessary until the initial condition is close to the equilibrium. In Fig. \ref{fig2}, the oscillations are significantly reduced for spin-up 2, and there are limited changes in the initial profiles and time evolution between spin-ups 3 and 4, suggesting the proper initial velocity field has been reached. It is also important to note that the wind speed and scaling parameters slowly increase over time in Figs. \ref{fig2}b-d in response to the growth of the CBL through entrainment.

One consequence of the long spin-up trials is that a low-level jet can develop within the entrainment layer directly above the heat flux minimum, particularly for weaker convection. This feature is not apparent from $U_x$ and $U_y$ for the profiles at $tf=\pi$ in the inset of Fig. \ref{fig2}a (yellow lines), and is more evident from the horizontal wind magnitude $U$. The jet is excluded from the rescaled profiles by assuming a linear trend in the top 100 m of $U_x$ and $U_y$ as seen in Fig. \ref{fig2}a. The jet is far from the region of interest below the mixed layer and understanding its emergence is outside the scope of the present study.

While Fig. \ref{fig2} is predominately for demonstration purposes, the same spin-up procedure was employed for the cases in Table \ref{tab1} to approximately determine the quasi-equilibrium wind profiles. For each simulated condition, four spin-up trials were completed on a grid with 200$\times$200$\times$100 points: the first with initial conditions $U_x=U_g$ and $U_y=0$, and the subsequent trials using the rescaled profiles as described above. The rescaled profiles resulting from the fourth trial were then used as initial conditions for the final simulations. It may be possible to expedite the determination of quasi-equilibrium conditions using existing methods not employed here. These options include initializing $U_x$ and $U_y$ based on an approximation of the cross-isobar angle and generating profiles using a single-column model \citep{Ghan2000}. Inertial oscillations in the latter option can be reduced by separating the equations based on the geostrophic ``background'' flow and the boundary layer deviations \citep{Gutman1973}.

The final simulations were spun-up on a larger grid with 400$\times$400$\times$200 points for approximately $t w_* / z_i \approx$ 15 turnover times. This duration ranges from 140 to 180 physical minutes for the different cases and is long enough for the flux profiles to develop, but short enough to avoid interference of the damping layer on the growing CBL. Because the initial $U_x$ and $U_y$ profiles are close to the equilibrium in the final spin-up, the inertial oscillations are minimized and it is not necessary to simulate a full inertial period.

After the spin-up, the velocity and temperature fields were interpolated onto the final grid with 1024$\times$1024$\times$512 points, and the simulations were continued for an additional 70 physical minutes. A short period is required for small-scale turbulence to develop within the finer grid, such that the first 10 minutes are excluded from the analysis. The time-averaged statistics for each case were computed across the last 60 minutes of the simulations. The time-averaged statistics are featured in Fig. \ref{fig1} and all later results.

\subsection{Grid sensitivity of near-surface statistics}
\label{subsec:converge}

For wall-modeled LES, flow statistics near the surface can be significantly biased by the wall model and grid resolution. To avoid inclusion of such biases in the analysis, the present section assesses the specific range of heights near the surface where the mean and gradient statistics are approximately converged with respect to grid resolution. More general discussions of grid resolution effects and mesh sensitivity are available elsewhere \citep[e.g.,][]{Davidson2009,Sullivan2011,Berg2020,Wurps2020}.

To test the sensitivity of results to grid resolution, cases A, C, and E from Table \ref{tab1} were repeated for the series of grid sizes summarized in Table \ref{tab2}. For a given case, all resolutions used the same initial velocity profiles and spin-up as determined from Sect. \ref{subsec:spinup}, with the different resolution introduced for the final 70 minutes of simulation. The sensitivity analysis only directly uses the two finest grids, but the additional coarser grids are useful for identify general trends discussed later in the context of the results.

\begin{table}
\centering
\caption{Numerical grids used to test the resolution convergence of flow statistics for cases A, C, and E in table \ref{tab1}. Grid properties include the number of nodes $N_j$ along each direction $j=$ $x$, $y$, and $z$, the domain size $L_j$, corresponding grid resolution $\Delta_j$, and effective resolution $\Delta = (\Delta_x \Delta_y \Delta_z)^{1/3}$ relative to the inversion height for case A. Results reported in the later results are based on the finest resolution.}
\label{tab2}       
\begin{tabular}{cccc}
\hline\noalign{\smallskip}
$N_x \times N_y \times N_z$	& $L_x \times L_y \times L_z$	& $\Delta_x \times \Delta_y \times \Delta_z$	& $\Delta/z_i$	\\
 (--)	& (km)	&(m)		& (--)	\\
\noalign{\smallskip}\hline\noalign{\smallskip}
100 $\times$ 100 $\times$ 50		& 12 $\times$ 12 $\times$ 2	& 120 $\times$ 120 $\times$ 40	& 0.080		\\
200 $\times$ 200 $\times$ 100	& 12 $\times$ 12 $\times$ 2	& 60 $\times$ 60 $\times$ 20		& 0.040		\\
400 $\times$ 400 $\times$ 200	& 12 $\times$ 12 $\times$ 2	& 30 $\times$ 30 $\times$ 10		& 0.020		\\
600 $\times$ 600 $\times$ 300	& 12 $\times$ 12 $\times$ 2	& 20 $\times$ 20 $\times$ 6.7	& 0.013		\\
800 $\times$ 400 $\times$ 400	& 12 $\times$ 12 $\times$ 2	& 15 $\times$ 15 $\times$ 5		& 0.010		\\
1024 $\times$ 1024 $\times$ 512	& 12 $\times$ 12 $\times$ 2	& 12 $\times$ 12 $\times$ 3.9	& 0.0078		\\	
\noalign{\smallskip}\hline
\end{tabular}
\end{table}

\begin{figure}
\centering
  \includegraphics[width=\linewidth]{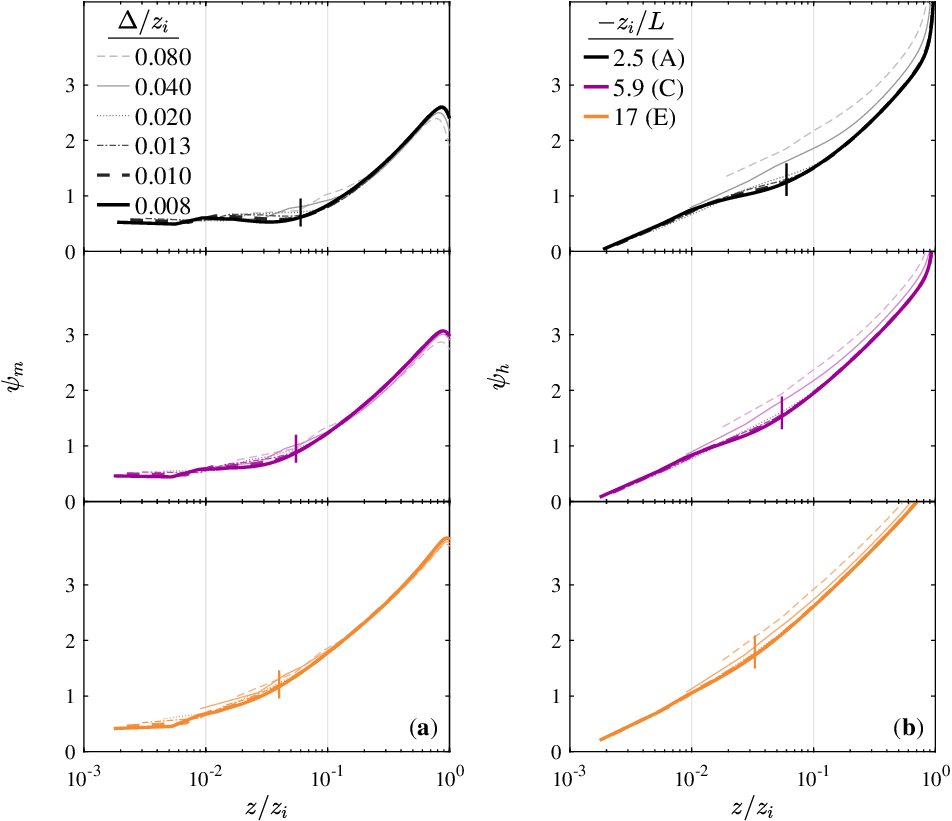}
\caption{Nondimensional mean wind speed (a) and air temperature (b) with varying grid resolution for cases A (top), C (middle) and E (bottom). Means are expressed as the diabatic term $\psi$ defined in Eq. \eqref{eq4}. Short vertical lines indicate the start of the converged region where the change in $\psi$ is less than 1\% between the finest two grids. The means are normalized using the von K\'{a}rm\'{a}n constant $\kappa=$ 0.39.}
\label{fig3}     
\end{figure}

Mean profiles of wind speed and air temperature are shown in Fig. \ref{fig3} for all tested grid sizes, with a vertical discplacement between different cases for visualization. The profiles are plotted as the diabatic term \citep{Panofsky1963}

\begin{align}
\psi_m &= \log{\left( \frac{z}{z_o} \right)} - \kappa \frac{U}{u_*} = \int\displaylimits_{-z_o/L}^{-\zeta} \frac{1 - \phi_m(\zeta^\prime)}{\zeta^\prime} d\zeta^\prime \nonumber\\
\psi_h &= \log{\left( \frac{z}{z_o} \right)} - \kappa \frac{\theta-\theta_s}{\theta_*} = \int\displaylimits_{-z_o/L}^{-\zeta} \frac{1 - \phi_h(\zeta^\prime)}{\zeta^\prime} d\zeta^\prime \label{eq4}
\end{align}

\noindent that quantifies the difference between the local mean value and the log law, where $\psi$ increases with convection. The value used here for the von K\'{a}rm\'{a}n constant is $\kappa=$ 0.39 \citep{Marusic2013}.

To more directly compare the profiles across resolution, the normalization in Fig. \ref{fig3} uses fixed scaling parameters $u_*$ and $\theta_*$ from the finest grid size rather than the parameter values resulting from each individual resolution. While there is an order of magnitude increase in resolution from the coarsest to finest grids in Table \ref{tab2}, the corresponding increase in $u_*$ is only 6.0\% for case A and 2.6\% for case E. The difference in $u_*$ between the two largest grids is 0.14\% for case A and 0.15\% for case E.

The criterion used to determine the converged region of the mean profiles is the percent difference in $\psi$ between the grids with 800$\times$800$\times$400 points and 1024$\times$1024$\times$512. Specifically, the difference in $\psi$ must be less than 1\% for this 28\% increase in resolution. The vertical lines in Fig. \ref{fig3} indicate the lowest height where this criterion is met, which varies from 0.04$z_i$ (case E) to 0.06$z_i$ (case A). The converged regions for cases B and D were inferred through interpolation, and the lower height in meters for case E was directly used for cases E1 and F. Later figures either exclude statistics in the near-surface region where convergence is not observed or clearly differentiate the unconverged results. Finally, while it is not apparent from Fig. \ref{fig3}, the mean wind speed and air temperature in the mixed layer are converged for the 600$\times$600$\times$300 grid in case A, and at coarser resolutions for stronger convection.

\begin{figure}
\centering
  \includegraphics[width=\linewidth]{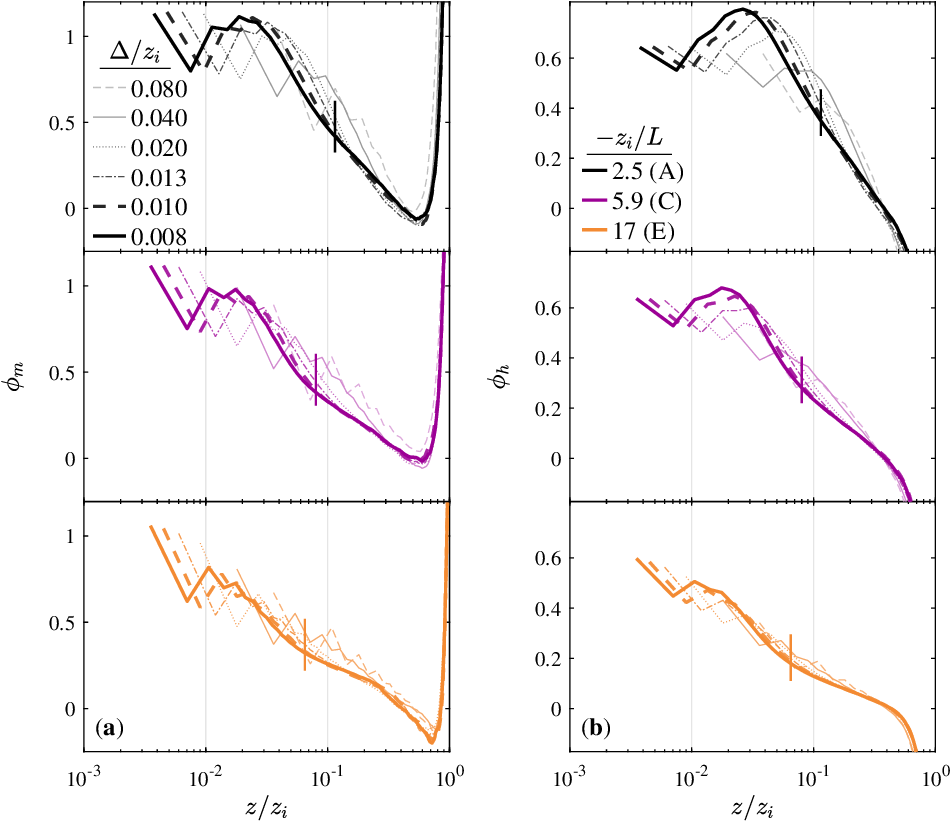}
\caption{Nondimensional gradients of velocity (a) and air temperature (b) with varying grid resolution for cases A (top), C (middle) and E (bottom). Gradients are expressed as $\phi$ defined in Eq. \eqref{eq1}. Short vertical lines indicate the start of the converged region where the change in $\phi$ is less than 0.01 between the finest two grids.}
\label{fig4}     
\end{figure}

A similar assessment is made for the nondimensional gradient profiles $\phi(z)$ in Fig. \ref{fig4}. The general concave shape of the profiles at the surface matches closely with previous observations \citep[e.g.,][]{BouZeid2005,Maronga2017} and demonstrates that the influence of the wall model, SGS model, and near-surface resolution extends well beyond the first grid points. The criterion used here for convergence of the gradient statistics is that the difference in $\phi$ must be less than 0.01 between the grids with 800$\times$800$\times$400 points and 1024$\times$1024$\times$512. Using the same 1\% threshold as above was found to be overly conservative given the very small magnitude of the derivatives far from the surface and would result in the defined converged regions beginning at moderately higher positions. The trends beginning near 0.8$z_i$ in Figs. \ref{fig3} and \ref{fig4} correspond to the start of the entrainment layer that is outside the scope of the present similarity analysis.

Under weak convection in case A, the gradient statistics do not converge within the traditional surface layer below 0.1$z_i$. There is also some uncertainty in the convergence across the lower half of the CBL, as the slope in $\phi$ appears to decrease with increasing resolution across the region with positive gradient. In moderate convection, both the upper portion of the surface layer and the outer layer exhibit grid convergence for the relatively fine resolution employed here. Figure \ref{fig4} demonstrates the challenge in using wall-modeled LES to critically evaluate near-surface behavior with a high degree of certainty. Accordingly, the conclusions drawn in the present work are confined to robust trends that extend beyond the surface layer and across all cases. The general findings are not contingent on the weakly convective cases that may not have reached mesh independence above the surface layer. With the threshold of 0.01, the converged regions of $\phi$ begin approximately 20 to 30 points away from the surface. The exact number of points is expected to depend on numerous variables including the LES wall and SGS models, grid size, grid aspect ratio, and convergence criterion, such that the quantitative outcomes of Figs. \ref{fig3} and \ref{fig4} are considered to be specific to this study.

\section{Mean velocity and temperature similarity}
\label{sec:results}

The profiles in Fig. \ref{fig1}, generated using the spin-up procedure outlined in Sect. \ref{subsec:spinup}, exhibit convergence with respect to grid resolution in the upper portion of the surface layer as discussed in Sect. \ref{subsec:converge}. Similarity in the region with converged statistics, including in the convective matching layer above the surface layer, is now evaluated in further detail. The upper half of the CBL is excluded from the results due to weak top-down effects from the entrainment layer that yield negative $\phi$ values as seen in Fig. \ref{fig4}.

\subsection{Comparison with Businger-Dyer relations}

The diabatic mean wind speed and air temperature profiles predicted by the Businger-Dyer relations in Eq. \eqref{eq3} result directly from the integration in Eq. \eqref{eq4} \citep{Paulson1970}:

\begin{align}
\psi_m &= 2\log{\left(\frac{1+x}{2}\right)} + \log{\left(\frac{1+x^2}{2}\right)} -2 \tan^{-1}x + \frac{\pi}{2} \nonumber\\
\psi_h &= 2\log{\left(\frac{1+x^2}{2}\right)}. \label{eq5}
\end{align}

\begin{figure}
\centering
  \includegraphics[width=\linewidth]{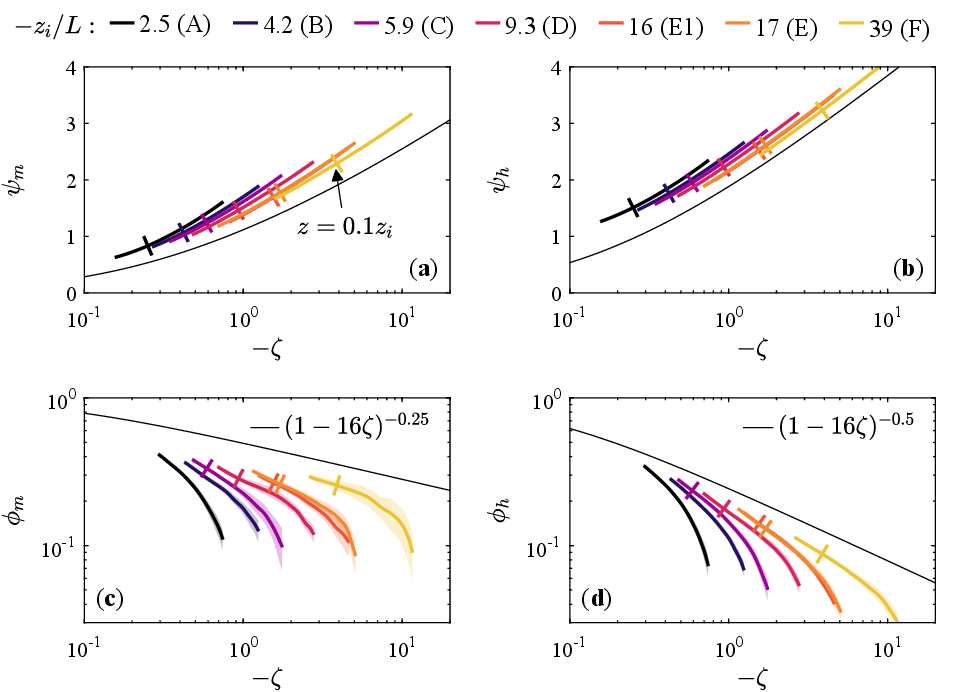}
\caption{Nondimensional profiles as a function of $\zeta = z/L$, compared to Businger-Dyer relations (thin lines) from Eqs. \eqref{eq3} and \eqref{eq5}. Rows correspond to the mean $\psi$ (a,b) computed from Eq. \eqref{eq4} and gradient $\phi$ (c,d) computed from Eq. \eqref{eq1}, and columns correspond to momentum (a,c) and heat (b,d). Shaded regions in (c,d) represent 95\% confidence intervals for statistical uncertainty in the mean gradients. Each profile includes heights from the bottom of the converged region determined in Sect. \ref{subsec:converge} up to 0.3$z_i$, and short lines indicate the height $z=0.1z_i$ for reference.}
\label{fig5}     
\end{figure}

\noindent Here, $x=\phi_m^{-1}=(1-16\zeta)^{0.25}$ and the contribution $\psi(-z_o/L)$ from the lower limit of the integral is neglected under the condition $-z_o/L \ll 1$. In Figs. \ref{fig5}a and \ref{fig5}b, the relations in Eq. \eqref{eq5} are compared with $\psi$ computed directly from the LES profiles using Eq. \eqref{eq4}. The range of heights included for each profile in Figs. \ref{fig5}a and \ref{fig5}b spans from the bottom of the converged region identified from Fig. \ref{fig3} up to 0.3$z_i$. The upper limit extends beyond the traditional definition for the surface layer and is included for consistency with the later analysis. The height 0.1$z_i$ is indicated in all figures where necessary to facilitate the distinction of trends within and above the traditional surface layer. While Eq. \eqref{eq5} appears to conform to the general shape of each LES profile, there is a distinct stability trend in which the profiles become increasingly offset from the prediction with weakening convection.

The displacement between cases is amplified in the gradient $\phi$ profiles shown in Figs. \ref{fig5}c and \ref{fig5}d. Consistent with previous observations, the general trend across cases appears to follow a curve similar to the Businger-Dyer profiles, but the decay in $\phi$ along each individual profile is significantly steeper \citep{Maronga2017,Pirozzoli2017,Li2018}. Unlike the referenced studies, however, the entirety of each $\phi$ curve is below the Businger-Dyer profiles. This difference is likely due to a combination of excluding the near-surface points from the present cases and the significant effect of inertial oscillations on the relevant normalization parameters $u_*$ and $L$ as seen in Fig. \ref{fig2}.

The shaded regions in Figs. \ref{fig5}c and \ref{fig5}d represent 95\% confidence intervals for the mean gradient statistics. The intervals are estimated from the formula $t_{95} \sigma_\phi / \sqrt{N}$, where $\sigma_\phi$ is the standard deviation of observed gradients over the 60-minute simulation period, $N$ is the number of independent observations, and $t_{95}$ is the Student's $t$ value that depends on the confidence level and $N$. The independent observations $N$ vary between 10 (moderate convection) and 20 (weak convection) across cases based on integral time scales computed from turbulent spectra. The steep decay in $\phi$ noted above and the separation of the curves in Fig. \ref{fig5} both exceed the statistical uncertainty.

Figure \ref{fig5} demonstrates that MOST and $\zeta$ alone cannot fully account for the LES profile trends. Scaling adjustments to the definitions of $\zeta$ and/or $\phi$ are required for the profiles within the surface layer to collapse along a common curve within the uncertainty bounds. At the same time, alignment with Businger-Dyer relations in field experiments and across different cases in simulation studies suggest that Eq. \eqref{eq3} provides a reasonably accurate foundation for similarity in the mean profiles. In the following section, the trends in $\phi$ are considered in the context of the extended profile to identify a possible similarity framework that is compatible with these past and present findings.

\subsection{Gradient profile trends}
\label{subsec:trends}

\begin{figure}
\centering
  \includegraphics[width=\linewidth]{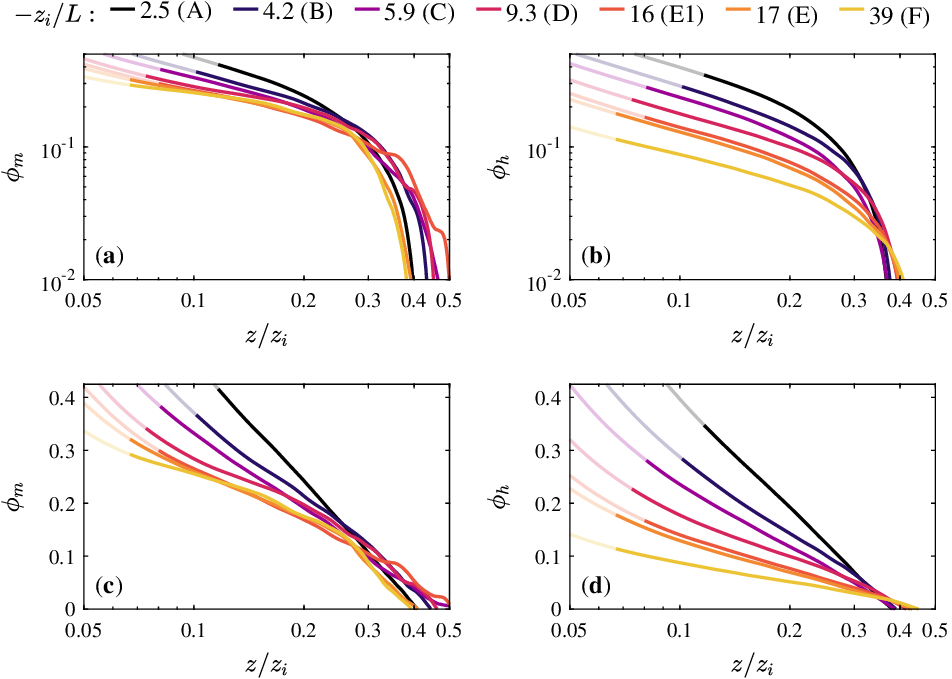}
\caption{Nondimensional gradient profiles as a function of $z/z_i$. Rows correspond to logarithmic (a,b) and log-linear (c,d) axis scaling, and columns correspond to momentum $\phi_m$ (a,c) and heat $\phi_h$ (b,d). Transparent regions of each curve are considered unconverged for the present resolution based on the assessment in Fig. \ref{fig4}.}
\label{fig6}     
\end{figure}

The nondimensional $\phi$ profiles in Figs. \ref{fig5}c and \ref{fig5}d are replotted in Fig. \ref{fig6} as a function of relative position $z/z_i$ within the CBL. The axes are shown with both logarithmic (top) and log-linear (bottom) scaling to facilitate interpretation of the profile shapes in different regions. The transparent extension of each curve is the region that did not meet the resolution convergence criterion detailed in Sect. \ref{subsec:converge} and Fig. \ref{fig4}.

The velocity gradients approaching the mixed layer exhibit incomplete statistical convergence, leading to modest random error along the profiles. Recalling that the observed values $\phi \lesssim O$(0.1) are the product of the gradient and the height $z\sim O$(100 m) in Eq. \eqref{eq1}, the dimensional gradient is exceedingly small in this region. Additional computing resources are not currently available to continue the simulations for a longer duration. Conclusions made from the gradient statistics are limited to trends that exceed the observed variability within each profile.

For the region near and below 0.1$z_i$ in Figs. \ref{fig6}a and \ref{fig6}b, there is noticeable curvature in the profile for weak convection in case A, but $\phi(z)$ increasingly resembles a power law with increasing convection. Further, the slope of the profiles, i.e. the exponent of the power law, does not vary across the tested cases. These trends are consistent with the form of the Businger-Dyer profiles in Eq. \eqref{eq3}, where the contribution of 1 vanishes with increasing convection and the power law exponent is assumed constant. With respect to grid resolution, the approximate power law in $\phi(z)$ under moderate convection is only apparent for the two largest grids tested in Table \ref{tab2}.

At higher positions, the mean gradients in Figs. \ref{fig6}a and \ref{fig6}b appear to be governed by an sharp decay in $\phi$ that results in the cutoff $\phi(z{=}0.4z_i)\approx 0$. The position of the cutoff is approximately constant as a fraction of $z_i$, whereas the start of the cutoff seen in Figs. \ref{fig5}c and \ref{fig5}d unambiguously varies with $L$. This has important implications for the approach to free convection and $-\zeta \rightarrow \infty$: rather than the mixed layer starting at lower positions and free convection reaching closer to the surface with decreasing $L$, Fig. \ref{fig6} suggests the near-surface region maintains a fixed height and $\phi$ decreases with increasing convection until it vanishes in the free convection limit.

The shape of $\phi$ approaching the mixed layer in Figs. \ref{fig6}c and \ref{fig6}d is approximately linear, particularly for the temperature. The linear trend implies a logarithmic decay in $\phi$ and a dimensional gradient that resembles $-\log(z)/z$. The height where the logarithmic decay gains leading order importance over the approximate power law appears to depend on stability and emerges at lower heights for the weaker convection cases. Importantly, the decay observed here is specific to a CBL with a well-defined mixed layer as seen in Fig. \ref{fig1}. For the tested grid sizes in Table \ref{tab2}, the logarithmic trend begins to appear with an overestimated slope for the grid with 400$\times$400$\times$200 points, and the slope is approximately converged for the grid with 600$\times$600$\times$300 points.

The trends in Fig. \ref{fig6}, i.e. the consistency with Eq. \eqref{eq3} closer to the surface and the logarithmic decay approaching the mixed layer height, show the potential to augment the existing Businger-Dyer profiles with an additional term that enforces the decay in $\phi$ with increasing $z/z_i$. In this sense, the term is a correction for the transition between the surface and mixed layers, where previous findings \citep[e.g.,][]{Salesky2012,Pirozzoli2017,Li2018} and Fig. \ref{fig5} indicate the correction is necessary even within the traditional surface layer below 0.1$z_i$. The existing model corrections for $\phi$ discussed in the introduction do not account for the specific trends observed in Fig. \ref{fig6}. For instance, the corrections based on $z_i/L$ displace $\phi$ by a constant value for a given case and do not consider the shape of the cutoff \citep{Cheng2021,Liu2022}. The cutoff is also not well described by an inverse summation of length scales \citep{Gryning2007}. A preliminary effort to empirically model the gradient cutoff is given in the following section.

\subsection{Preliminary model for extended similarity}

Deeper within the ABL in the roughness sublayer (RSL), the mean profiles are influenced by complex turbulent drag and mixing interactions associated with the local surface roughness. In this sublayer, it is standard to correct for the mean similarity in a multiplicative manner as $\phi_m(\zeta) \varphi_m(z/z_*)$, where $\phi_m$ accounts for atmospheric stability and $\varphi_m$ is a correction based on the relative position within the RSL depth $z_*$ \citep[e.g.,][]{Garratt1980,Cellier1992,Molder1999}. The most common functional form for $\varphi$ is an exponential that models the decay in the gradient within the RSL as the surface is approached \citep{Garratt1980,Harman2007,Mo2022}. The same form is adopted here, except the purpose of the exponential in this case is to ensure the gradient decreases towards zero approaching the mixed layer rather than within the RSL. \citet{Santoso1998,Santoso2001} also used an exponential cutoff correction to the mean profiles to enforce a smooth transition to uniform mixed layer values. The revised similarity relations are given as

\begin{align}
\phi_m(\zeta,z/z_i) &= (1- b_m \zeta)^{-0.25} \exp{\left( -c_m \frac{z}{z_i} \right)} \nonumber \\
\phi_h(\zeta,z/z_i) &= a_h (1 - b_h \zeta)^{-0.5} \exp{\left( -c_h \frac{z}{z_i} \right)} , \label{eq6}
\end{align}

\noindent where $a$, $b$, and $c$ are fitted constants and the leading constant $a_h$ for temperature accounts for the turbulent Prandtl number. The exponent values in Eq. \eqref{eq3} are adopted here, noting that testing a wide range of exponents resulted in values within $\pm0.05$ of 0.25 and 0.5 and only a nominal increase in the coefficient of determination $R^2$ for the fit. The exponential is included in the definition of $\phi$ in Eq. \eqref{eq6} rather than as a separate $\varphi$ function because it is part of the same stability correction. The constant $c$ determines the height relative to $z_i$ where the exponential becomes small. A value $c>2$ is expected such that the cutoff function decreases to small values within the lower half of the CBL.

To evaluate the applicability of the revised similarity relations, the expressions for $\phi$ in Eq. \eqref{eq6} and their integral $\psi$ defined in Eq. \eqref{eq4} were fitted to the LES profiles. The integral for $\psi$ was computed numerically in the absence of a simple analytical solution. The reason for including $\psi$ in the fitting procedure is to assess the extension of $\phi$ down to $z=z_o$. While the near-surface region cannot be fitted directly, Eq. \eqref{eq6} must have the correct cumulative magnitude below the fitted region in order to align with the LES profiles for $\psi$.

The cost function for the nonlinear fitting algorithm was the total residual between the predicted and observed $\phi$ and $\psi$ values compiled for all cases simultaneously. The fit result therefore represents the range of convective conditions rather than any individual case. The fitting procedure was conducted separately for the velocity and temperature statistics. Due to the complexity of the equations and the use of numerical integration, the algorithm was unable to converge when multiple parameters were undefined. Accordingly, the fit was designed to optimize $c$ with $a$ and $b$ as prescribed inputs, and was repeated for a range of $a$ and $b$ values. The values presented here are those with the highest resulting $R^2$, with $a_m=1$ assumed for velocity. As noted above, the power law exponents in Eq. \eqref{eq6} were also varied before the traditional values were selected for simplicity. Finally, heights up to 0.3$z_i$ were included in the fits under the assumption that Eq. \eqref{eq6} approximates the transition across an extended range up to the convective mixed layer.

The preliminary values resulting from the fit to the velocity profiles are $b_m \approx 22$, $c_m \approx 3.7$, and $R^2=0.974$. The values for the temperature profiles are $a_h \approx 0.93$, $b_h\approx 14$, $c_h\approx 2.9$, and $R^2=0.992$. The higher $R^2$ for temperature is likely due in part to the additional fitted parameter $a_h$ and the better statistical convergence of $\phi$ in Fig. \ref{fig6} relative to velocity. Owing to the lack of near-surface points, the present values for $a_h$ and $b$ are not suggested as replacements to existing values. Further, the difference between $c_m \approx 3.7$ and $c_h \approx 2.9$ is not given a physical interpretation here. It may result simply from the fact that the smaller Businger-Dyer exponent $-0.25$ for momentum requires a larger cutoff correction to have similar $\phi$ near the mixed layer. The main outcome of the fit is to demonstrate the systematic improvement of the profile prediction with the inclusion of a $z/z_i$ cutoff.

\begin{figure}
\centering
  \includegraphics[width=\linewidth]{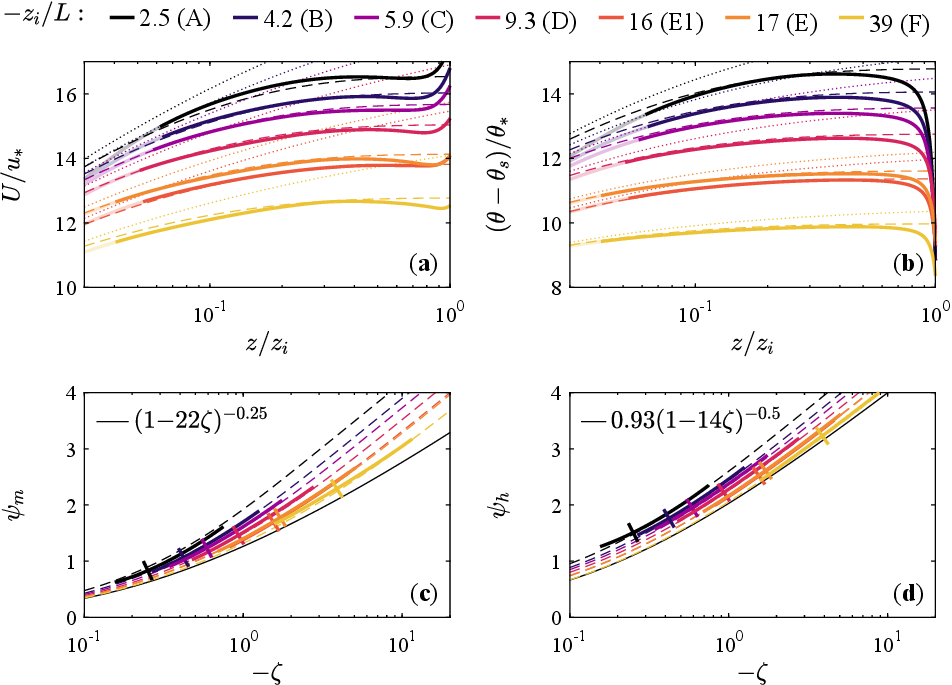}
\caption{Comparison of LES mean profiles (thick lines) with the best fit of the Eq. \eqref{eq6} integral (dashed lines) and the same equation without the exponential cutoff (dotted lines). Rows correspond to the mean value (a,b) and the diabatic term $\psi$ (c,d), and columns correspond to momentum (a,c) and heat (b,d). Transparent regions of the curves in (a,b) are considered unconverged for the present resolution based on the assessment in Fig. \ref{fig3}. Each profile in (c,d) includes heights from the bottom of the converged region up to 0.3$z_i$, and short lines indicate the height $z=0.1z_i$ for reference.}
\label{fig7}     
\end{figure}

Figure \ref{fig7} compares the mean profiles predicted from the integral of Eq. \eqref{eq6} (dashed lines) with the LES profiles. The dotted lines in Figs. \ref{fig7}a and \ref{fig7}b are Eq. \eqref{eq6} with the fitted values, but excluding the exponential cutoff. When plotted as $\psi$, these dotted lines collapse along the solid black lines in Figs. \ref{fig7}c and \ref{fig7}d defined by the integral of the equations given in the legend.

The expression with the exponential cutoff leads to significant improvements in the alignment with the LES profiles in Figs. \ref{fig7}a and \ref{fig7}b across all heights within the converged region, particularly for heights above 0.1$z_i$. While there are some discrepancies, most notably for cases A and F, the predictions resulting from Eq. \eqref{eq6} provide a reasonable approximation of the mean profiles for an extended range from below 0.1$z_i$ to above 0.3$z_i$ and near the start of the convective mixed layer.

The diabatic term $\psi$ in Figs. \ref{fig7}c and \ref{fig7}d demonstrates the departure of the mean profiles from surface layer similarity as a result of the decay in the gradients along $z/z_i$. The dashed lines representing Eq. \eqref{eq6} all begin at $\psi(z{=}z_o)=0$ and become increasingly dissimilar from Monin--Obukhov scaling similarity with increasing $z/z_i$, i.e. the $\psi$ values spread farther apart. This dissimilarity is well predicted by the exponential cutoff correction.

\begin{figure}
\centering
  \includegraphics[width=\linewidth]{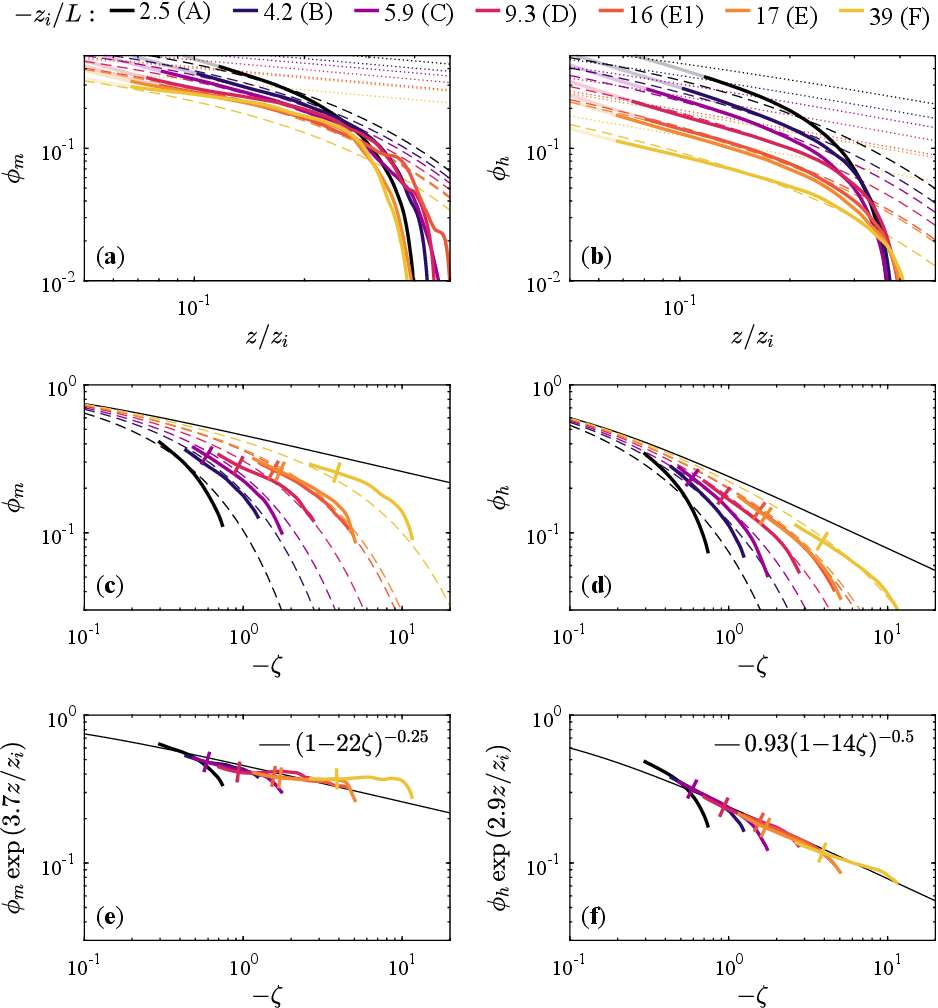}
\caption{Comparison of LES $\phi$ profiles (thick lines) with the best fit of Eq. \eqref{eq6} (dashed lines) and the same equation without the exponential cutoff (dotted lines). Rows correspond to the profiles plotted versus $z/z_i$ (a,b), plotted versus $\zeta$ (c,d), and after compensating for the exponential cutoff (e,f). Columns correspond to momentum (a,c,e) and heat (b,d,f). Transparent regions of the curves in (a,b) are considered unconverged for the present resolution based on the assessment in Fig. \ref{fig4}. Each profile in (c-f) includes heights from the bottom of the converged region up to 0.3$z_i$, and short lines indicate the height $z=0.1z_i$ for reference.}
\label{fig8}     
\end{figure}

Figure \ref{fig8} evaluates the gradient profiles predicted from Eq. \eqref{eq6} (dashed lines) in the same manner as the mean values in Fig. \ref{fig7}. As before, the dotted lines in Figs. \ref{fig8}a and \ref{fig8}b exclude the exponential cutoff and are equivalent to the solid black lines in Figs. \ref{fig8}c-f defined by the equations in the legends.

Equation \eqref{eq6} matches closely with the $\phi$ curves in Figs. \ref{fig8}a and \ref{fig8}b compared to the Businger-Dyer profiles without a cutoff correction. However, the exponential cutoff does not fully account for the entirety of the gradient decay, as seen in the deviations from the LES profiles that emerge between 0.2-0.3$z_i$. As discussed previously, the dimensional value of the gradients in this range of heights is very small such that the discrepancy may be of limited practical importance. For instance, the discrepancy in $\phi$ approaching the mixed layer is not readily seen in the Figs. \ref{fig7}a and \ref{fig7}b mean profiles.

The plots of $\phi(\zeta)$ in Figs. \ref{fig8}c and \ref{fig8}d show the departure from Monin--Obukhov similarity in the LES profiles compared with the prediction from the exponential cutoff. The exponential cutoff closely approximates the decay in the gradients within and above the surface layer for the fitted cases. To collapse the nondimensional gradients along a single curve, it is necessary to group the gradient and cutoff correction as $\phi \exp{(cz/z_i)}$. The product, shown in Figs. \ref{fig8}e and \ref{fig8}f, now aligns reasonably well with the Businger-Dyer profiles. Most of the residual differences occur near 0.3$z_i$ and are due to the discrepancies seen in Figs. \ref{fig8}a and \ref{fig8}b and discussed above.

Figure \ref{fig8} provides promising evidence for expanding the similarity parameter space to include $z/z_i$. The correction in Eq. \eqref{eq6} provides profile predictions for an extended range approaching the mixed layer and may account for the departures from MOST observed in previous simulation studies \citep{Khanna1997,Pirozzoli2017,Li2018}. However, the generality of the results remain unproven at this point and further comparison with other simulations and measurements is required. In particular, the exponential form of the correction may be specific to the use of the Businger-Dyer profiles to functionally represent MOST. Alternate empirical relations with a steeper decay for strong convection with large $\zeta$ would require a different correction. However, the necessity to expand the parameter space to achieve universality of profiles in Fig. \ref{fig8} is independent of the empirical MOST function used.

If Eq. \eqref{eq6} is applicable beyond the present cases, one interesting note is that the cutoff correction is independent of stability. The LES cases in Table \ref{tab1} span the transition from relatively weak convection with thermal rolls to moderately strong convection with cells \citep{Etling1993,Atkinson1996,Khanna1998,Salesky2017}. For the current results, the correction at a given height $z/z_i$ is the same regardless of the convective regime, indicating that the different large-scale structures (i.e. rolls or cells) impinging on the surface layer reduce the average gradient by the same fraction.

\section{Discussion}
\label{sec:discussion}

\subsection{Implications for mixed layer resistance dependencies}

While there is extensive theory for resistance laws in the geostrophic drag and heat transfer across the entire ABL \citep[see, e.g.][]{Monin1970,Yamada1976,Arya1977}, there are relatively fewer studies relating mean wind speed $U_m$ and temperature $\theta_m$ in the convective mixed layer to surface properties. The predicted scaling for $U_m/u_*$ and $(\theta_m-\theta_s)/\theta_*$ depends on underlying assumptions, in particular regarding the bottom height $z_m$ of the mixed layer. The dependencies of the mean values can be inferred by evaluating the mean profiles at the base of the mixed layer $z=z_m$:

\begin{align}
\frac{U_m}{u_*} &= \frac{1}{\kappa} \left[ \log{\left( \frac{z_m}{z_o} \right)} - \psi_m\left( \frac{z_m}{L},\frac{z_m}{z_i}\right) \right] \nonumber\\
\frac{\theta_m-\theta_s}{\theta_*} &= \frac{1}{\kappa} \left[ \log{\left( \frac{z_m}{z_o} \right)} - \psi_h\left( \frac{z_m}{L},\frac{z_m}{z_i}\right) \right]. \label{eq7}
\end{align}

\noindent Equation \eqref{eq7} is a direct outcome of the definition for the diabatic term in Eq. \eqref{eq4}, noting that traditional approaches do not include the $z/z_i$ correction for $\psi$. If $z_m \propto z_i$ is assumed in Eq. \eqref{eq7}, the resulting mixed layer values depend directly on both $\log{(z_i/z_o)}$ and $\psi(z_i/L)$ \citep{Garratt1982}. Alternatively, if the $z/z_i$ correction is excluded and $z_m \propto -L$ is assumed based on arguments of local free convection \citep{Wyngaard1971}, the result in Eq. \eqref{eq7} depends solely on $\log{(-L/z_o)}$ \citep{Zilitinkevich1992,Liu2023}. This same dependency has been derived for a mixed layer velocity scale using matched asymptotic expansions with three scaling layers spanning the surface layer and no assumptions for $z_m$ \citep{Tong2020}.

\begin{figure}
\centering
  \includegraphics[width=\linewidth]{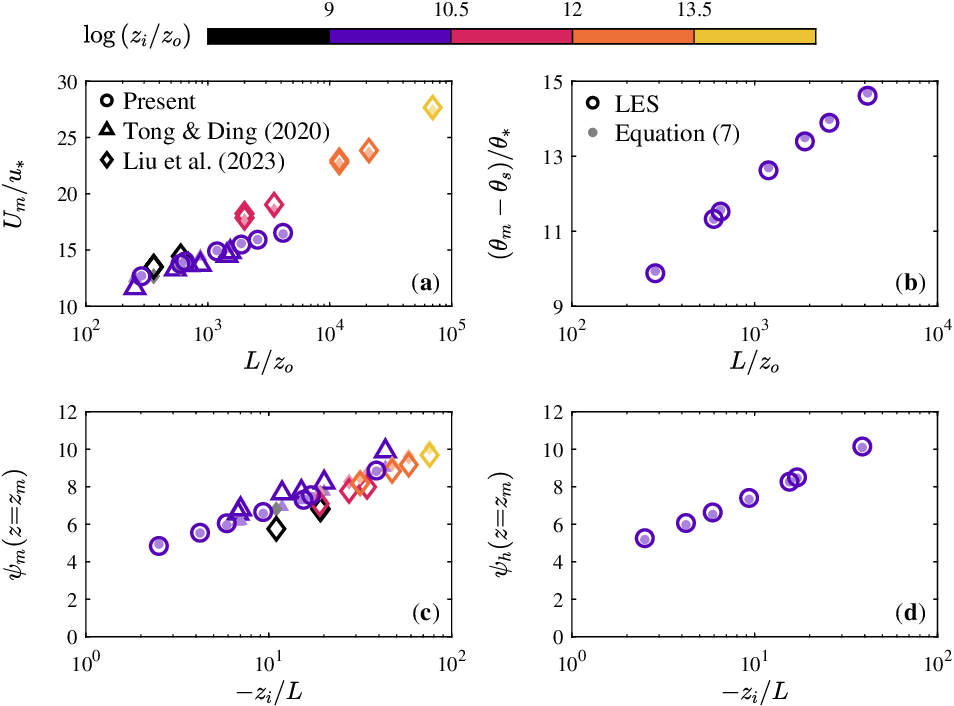}
\caption{Dependence of the mean velocity $U_m$ and temperature $\theta_m$ in the convective mixed layer. Rows correspond to the mean values versus $L/z_o$ (a,b) and their diabatic terms versus $z_i/L$ (c,d), and columns correspond to velocity (a,c) and temperature (b,d). Results are included for the present LES and two reference LES studies indicated by open symbols, where each are compared to predicted values based on Eq. \eqref{eq6} indicated by closed transparent symbols. Observed and predicted values are based on the mixed layer bottom height $z_m=0.4z_i$. Symbol color corresponds to $z_i/z_o$.}
\label{fig9}     
\end{figure}

The results in Sect. \ref{subsec:trends} and Fig. \ref{fig6} yield $z_m \approx 0.4z_i$ for all LES cases based on the logarithmic decay of the gradients. Considering Eq. \eqref{eq6} aligns with the mean profiles up to the mixed layer in Figs. \ref{fig7}a and \ref{fig7}b, the present findings indicate that the mixed layer values in Eq. \eqref{eq7} should depend on both $z_i/z_o$ and $z_i/L$ in a complex manner. To test this, the mixed layer mean values at $z_m=0.4z_i$ are shown in Fig. \ref{fig9}. The values in Fig. \ref{fig9} are insensitive to the choice of $z_m/z_i$ within the range 0.4$-$0.8 due to the uniformity of the mixed layer seen in Fig. \ref{fig1}. Because $U_m$ and $\theta_m$ are mean values, the purpose of Eq. \eqref{eq7} and Fig. \ref{fig9} is to assess the dependencies discussed above and not to define new velocity and temperature scales for the mixed layer. Included in Fig. \ref{fig9} for comparison are the predicted values from numerical integration of Eq. \eqref{eq6} up to $z_m$. It may be possible to evaluate further the integral definition for $\psi (z_m/L,z_m/z_i)$ \citep[see, e.g.,][]{Physick1995,deRidder2010}, but for the present discussion the primary concern is the $\log{(z_m/z_o)}$ term in Eq. \eqref{eq7} that is already uncoupled from $\psi$.

The velocity statistics in Fig. \ref{fig9} are supplemented with results from two recent LES studies \citep{Tong2020,Liu2023}. The \citet{Tong2020} points only represent their LES cases with the \citet{Kosovic1997} SGS model. Further, the $U_m$ values reported in \citet{Tong2020} are a velocity scale for the mixed layer that differs in value from the mean velocity (see, e.g., their Fig. 3), such that the $U_m$ values used in Fig. \ref{fig9} were inferred from their CBL profiles. The differing log-linear slope in Fig. \ref{fig9}a for the present LES and the cases from \citet{Liu2023} indicate that $\log{(-L/z_o)}$ is not the sole determinant of $U_m/u_*$. In particular, increasing $z_i/z_o$ leads to a vertical shift in the resistance value.

The predicted values (closed symbols) for the external references use the same parameter values fitted to the present LES. The combination of the corrected similarity expression in Eq. \eqref{eq6} and the fixed height $z_m=0.4z_i$ lead to an accurate prediction of the differing trends noted above for Fig. \ref{fig9}a. The alignment with the reference studies is promising regarding the potential generality of the exponential cutoff correction.

The roughness dependence leading to a vertical shift in the mixed layer resistance value can be offset by plotting the diabatic term $\psi_m$ as in Fig. \ref{fig9}c. In this format, the data are approximately aligned along a common curve, noting that the residual differences may be due in part to the spin-up procedure that can significantly affect $L$ (Fig. \ref{fig2}) and differences between SGS models as observed in \citet{Tong2020}. Importantly, $\psi$ will also vary with roughness as $z_o$ becomes large, but $z_o \ll z_m$ for these data and its contribution to $\psi$ is negligible.

Fig. \ref{fig9} supports $z_m \propto z_i$ and the dependence of $U_m/u_*$ on both $z_i/z_o$ and $z_i/L$. However, the combination of Eqs. \eqref{eq6} and \eqref{eq7} do not form a drag law and further work is required to develop the result into a velocity scale. Further, no conclusion can be made for temperature due to lack of independent data across a range of $z_i/z_o$. The mixed layer temperature resistance is included for the present LES in Figs. \ref{fig9}b and \ref{fig9}d for completeness.

\subsection{Implications for the surface layer height}

\begin{figure}
\centering
  \includegraphics[width=\linewidth]{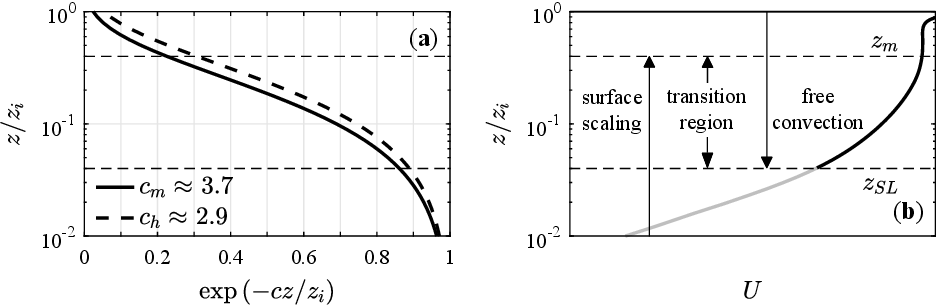}
\caption{Depiction of the transition region between the top of the surface layer $z_{SL}$ and the bottom of the mixed layer $z_m$. (a) Value of the exponential cutoff in the gradient profiles for momentum and heat. (b) Contribution of both surface scaling and mixed layer free convection events to the mean gradients within the transition region, where the relative contribution of each varies with height in conjunction with the cutoff function in (a).}
\label{fig10}     
\end{figure}

The extended logarithmic decay of $\phi$ as the gradients vanish to zero in Fig. \ref{fig6} provide a consistent criterion for defining the bottom of the mixed layer where free convection occurs, but defining the top of the surface layer $z_{SL}$ is more ambiguous. The $z/z_i$ cutoff correction in Eq. \ref{eq6} is non-negligible within the full range of heights analyzed here, from approximately 0.05$z_i$ up to $z_m$, where the correction is required to account for simulation trends observed in Fig. \ref{fig8} and in previous studies \citep[e.g.,][]{Khanna1997,Pirozzoli2017,Li2018}. Upon further considering the nonlocal contribution of large-scale eddies to the decay in the gradients at these heights \citep{Li2018,Fodor2019}, it is possible that the extended range from $z_{SL}$ (not yet defined) to $z_m$ is a transition region resulting from the coexistence of local and nonlocal eddies governed by different mechanisms and scales. Here, the exponential cutoff approximates the transition in the mean profiles from Monin--Obukhov similarity in the surface layer to the zero gradient condition characterizing the mixed layer. This transition is depicted in Fig. \ref{fig10}.

Assuming the exponential cutoff and fitted constant $c$ in Fig. \ref{fig10}a represents the appropriate correction to surface similarity, the correction is less than 10\% only for heights in the lowest few percent of the CBL depth. For $z_i \sim O$(1 km), these heights correspond to the lowest 30 m or so of the atmosphere where many field measurements are sampled. On one hand, this indicates that the correction is not significant for many previous field campaigns and explains why the original Businger-Dyer profiles align well with experimental data to within the uncertainty and scatter of the results. On the other hand, a 10\% correction at 0.03$z_i$ is non-negligible and emphasizes the need to reconsider the conventional height for $z_{SL}$.

Derivations for the generalized log law $U/u_* = \kappa^{-1} \left[ \log{(z/z_o)} - \psi_m(\zeta) \right]$ in the inertial sublayer of the ABL often rely on asymptotic matching between the surface conditions and the outer layer where the velocity defect follows Rossby-number similarity \citep[see, e.g.,][]{Blackadar1968,Hess1973}. These derivations use the assumption $z/z_i \rightarrow$ 0 (i.e. $z \ll z_i$) to define $z_{SL} \sim 0.1z_i$ and do not consider the extensive uniformly mixed layer in convective conditions. Owing to the influence of the mixed layer on near-surface statistics, a new assumption $z/z_m \rightarrow$ 0 for the surface layer depth is suggested here. While it may be possible to formalize a revised asymptotic matching between the surface conditions and the mixed layer, the matching is complicated by the dependence of the mixed layer resistance on surface properties as seen in Eq. \eqref{eq7}.

In Fig. \ref{fig10}, the approximation $z_{SL} \approx 0.1z_m  \approx 0.04z_i$ is applied under the assumption that $z_{SL} \ll z_m$ is required for the contribution of nonlocal eddies to the gradient -- and the resulting correction factor -- to be small. However, the correction is greater than 10\% at this height such that a more stringent definition may be warranted. A formal framework is required to address this issue of surface layer height in a more rigorous manner. Regardless of the exact definition for $z_{SL}$, the traditional estimate $z_{SL} = 0.1z_i$ is insufficient for convective conditions based on the growing body of evidence discussed in the introduction.

The question of defining $z_{SL}$ is complemented by recent evidence that mean profile similarity in the surface layer of the stable ABL also depends on the boundary layer depth \citep{Heisel2023}. While the flow structure for stable conditions is considerably different, the same general reasoning applies: z-less stratification above the surface layer indirectly influences the gradients below 0.1$z_i$. Here, free convection turbulence above the surface layer influences the gradients below 0.1$z_i$ in a more direct manner.

The transition region depicted in Fig. \ref{fig10}b coincides with the regime previously discussed as a local free convection layer \citep{Tennekes1970,Kader1990} or convection matching layer \citep{Panofsky1978,Kaimal1994}. However, formulations for those layers do not transition smoothly into the uniformly mixed layer \citep{Panofsky1978}, unlike the present work and similar studies of the so-called radix layer \citep{Santoso1998,Santoso2001}. There is also a distinction in the observed scaling, at least for the first-order statistics. In the traditional local convection layer with $-L \ll z \ll z_i$, the $z$ scaling is still relevant but the velocity and temperature variables no longer depend on $u_*$ \citep{Wyngaard1971}. In the present study, the exponential cutoff reduces the gradient based on the relative position within the transition region, but does not alter the scaling for $\phi$ or $\zeta$. As an analogy, the fluxes $-\overline{u^\prime w^\prime}/u_*^2$ decay with $z/z_i$ but scale with $u_*$ throughout the CBL depth. In this sense, the gradients maintain their surface scaling up to $z_m$ despite the incomplete similarity due to the $z/z_i$ correction demonstrated in Fig. \ref{fig8}.

This $z/z_i$ correction corresponds to the presence of large-scale eddies such as downdrafts \citep{Li2018} and updrafts \citep{Fodor2019} whose governing scale $w_*$ is oriented along $z$ \citep{Deardorff1970}. Within the mixed layer, the ensemble of these predominately vertical motions results in zero mean gradient. If the free convection events that extend into the transition region also have a collective mean gradient close to 0, the events would contribute to a decrease in the overall $\phi$ without incurring a statistically meaningful transition in scaling from $u_*$ to $w_*$. The decrease in $\phi$ would then directly depend on the relative probability of free convection events that increases with height in a manner consistent with the exponential in Fig. \ref{fig10}a. Importantly, the same argument cannot be extended to the higher-order variance statistics and an analysis of the variances is outside the scope of the present work.

There is uncertainty in Fig. \ref{fig10} with respect to changes in the behavior with increasing instability. The present analysis suggests the exponential cutoff and parameter $c$ do not vary with $z_i/L$. As noted previously, this suggests the correction and the relative gradient contributions are independent of changes in the eddy topology from roll structures to vertical cells. Further, with $z_m \propto z_i$ the surface-scaling region in Fig. \ref{fig10}b must increasingly resemble the mixed layer turbulent structure in order for the surface layer to vanish in free convection. These implications should be evaluated across a wider range of $z_i/L$ and for a greater number of cases before conclusions are drawn.

\section{Summary}
\label{sec:conclusion}

The present work uses a series of seven LES cases to study mean profile similarity in the lower half of the convective boundary layer. The cases represent a dry, barotropic idealized CBL under weak to moderately strong convection with mid-latitude Coriolis frequency, a stable capping inversion, and a well-defined mixed layer. An ad hoc spin-up procedure is used to mitigate inertial oscillations in the final simulations (Fig. \ref{fig2}), and the grid converge of near-surface profile statistics is closely examined (Figs. \ref{fig3} and \ref{fig4}).

The new LES cases reveal the same qualitative trends in the nondimensional gradients $\phi$ seen in other recent simulation-based studies (Fig. \ref{fig5}): the results generally align with Monin--Obukhov similarity across the different cases, but the individual profiles each exhibit a steeper slope that precludes universal similarity in $\phi(\zeta)$ \citep{Pirozzoli2017,Maronga2017,Li2018}. In other words, MOST captures variability in $\phi$ across a range of $L$ and fixed $z$, but does not fully account for the variability across $z$ for fixed $L$. The behavior of the $\phi$ profiles above the surface layer indicates that the steeper slope is associated with a broader trend in $z/z_i$ that reduces the gradient towards 0 at the height of the mixed layer (Fig. \ref{fig6}).

To account for this trend, the well-known Businger-Dyer profiles are revised in Eq. \eqref{eq6} with an exponential cutoff similar to corrections for similarity in the roughness sublayer \citep{Garratt1980,Harman2007}. The revised expressions, with fitted parameters $c_m\approx$ 3.7 and $c_h\approx$ 2.9 for the exponential term $\exp{(-cz/z_i)}$, result in significantly improved similarity for the mean (Fig. \ref{fig7}) and gradient (Fig. \ref{fig8}) profiles from approximately 0.05$z_i$ up to the mixed layer near 0.4$z_i$. The correction is expected to be small close to the surface where most point measurements are acquired in field experiments, which may explain why the consistent $z/z_i$ trend seen in simulations is not readily apparent from field observations. Further, the parameter space probed by field measurements often spans a wide range of $L$ at a limited series of fixed heights, which as noted above can yield curves that closely follow MOST relations.

In addition to the improved similarity, there are three important implications arising from the revised relations in Eq. \eqref{eq6}. First, the mean values in the mixed layer depend on both $\log{(z_i/z_o)}$ and $\zeta$ (Fig. \ref{fig9}). Second, the exponential correction accounts for an extended transition region in the mean profiles between the surface layer and the mixed layer (Fig. \ref{fig10}), where this region is strongly influenced by large-scale buoyancy-driven eddies \citep{Li2018,Fodor2019}. Owing to the effect of these eddies, the correction only becomes small for $z/z_i \sim O$(0.01), such that the common assumption of 0.1$z_i$ for the surface layer height is too large for idealized convective conditions. Third, the start of the mixed layer at a fixed fraction of $z_i$ (Fig. \ref{fig6}) suggests the surface layer turbulent structure changes with increasing convection in order to match the mixed layer under free convection, but extending the analysis to stronger convection is required to validate the last point.

While the fitted results in Figs. \ref{fig7} and \ref{fig8} are promising, the present analysis includes a limited number of cases and lacks reliable statistics in the bottom half of the surface layer. In particular, modest grid dependence is observed across the surface layer for weak convection in case A (Fig. \ref{fig4}). Equation \eqref{eq6} is thus considered to be a preliminary effort to model the trends in $\phi$ observed in Figs. \ref{fig5} and \ref{fig6}. Several other functional forms were evaluated to more accurately account for the logarithmic decay in $\phi$, but the collective profile data were found to be prone to overfitting, where the functional dependencies were borne by the fitted parameters. Accordingly, the proposed correction is limited to an extension of the widely-tested Businger-Dyer profiles, and the correction has only one nondimensional parameter ($c_m$ or $c_h$) that has a clear physical interpretation corresponding to the height where the cutoff reaches a given magnitude. Importantly, the present quantitative correction depends on the empirical profiles being corrected, i.e. Businger-Dyer, O'KEYPS, or another alternative. However, the lack of universality in $\phi(\zeta)$ in Fig. 5 and the necessity to include $z/z_i$ in mean profile similarity are general findings that do not depend on the specific empirical relations. Further analysis with additional datasets may support a more sophisticated model that better matches the full gradient cutoff up to 0.4$z_i$ in Figs. \ref{fig8}a and \ref{fig8}b, but for the available data in the present study only a simpler model is warranted.

\section*{Declarations}

The authors have no competing interests to declare. Profiles of the LES mean and flux statistics are published in the Zenodo data repository: \url{https://doi.org/10.5281/zenodo.10414843}.

\begin{acknowledgements}
The authors gratefully acknowledge high-performance computing support from Cheyenne (doi:10.5065/D6RX99HX) provided by the National Center for Atmospheric Research Computational Information Systems Laboratory. Additionally, M. H. acknowledges start-up support from the School of Civil Engineering at the University of Sydney and M. C. acknowledges partial funding support from the Biological and Environmental Research program of the U.S. Department of Energy (DE-SC0022072).
\end{acknowledgements}

\bibliographystyle{spbasic_updated}     
\bibliography{references} 

\end{document}